\documentclass[preprint, onecolumn, showpacs, amsmath]{revtex4}
\usepackage{dcolumn}
\usepackage{bm}
\usepackage{graphicx}
\usepackage{color}
\DeclareGraphicsExtensions{. jpg,. pdf, . mps, . png, . eps, . ps, . EPS}

\begin{document}
\def\be{\begin{equation}}
\def\ee{\end{equation}}

\def\bc{\begin{center}} 
\def\ec{\end{center}}
\def\bea{\begin{eqnarray}}
\def\eea{\end{eqnarray}}
\newcommand{\avg}[1]{\langle{#1}\rangle}
\newcommand{\Avg}[1]{\left\langle{#1}\right\rangle}

\title{Enhancement of $T_c$ in the Superconductor-Insulator Phase Transition on Scale-Free Networks }

\author{Ginestra Bianconi}

\affiliation{Department of Physics, Northeastern University, Boston, 
Massachusetts 02115 USA.\\
 }

\begin{abstract} 

 A road map to understand the relation between  the onset of the superconducting state with the particular optimum heterogeneity  in granular superconductors is to study a Random Tranverse Ising Model on complex networks with a scale-free degree distribution regularized by and exponential cutoff $p(k)\propto k^{-\gamma}\exp[-k/\xi]$. In this paper we characterize in detail the phase diagram of this model and its critical indices both on annealed and quenched networks. To uncover the phase diagram of the model we use   the tools of heterogeneous mean-field calculations for the annealed networks  and the most advanced techniques of quantum cavity methods for the quenched networks. The phase diagram of the dynamical process depends  on the temperature $T$, the coupling constant $J$ and on  the value of the branching ratio $\frac{\avg{k(k-1)}}{\avg{k}}$ where $k$ is the degree of the nodes in the network. For fixed value of the coupling the critical  temperature increases linearly with $\frac{\avg{k(k-1)}}{\avg{k}}$ which diverges with the increasing cutoff value $\xi$ for value of the $\gamma$ exponent $\gamma\leq 3$. This result suggests that the fractal disorder of the superconducting material can be responsible for an enhancement of the superconducting critical temperature.
At low temperature and low couplings $T\ll1$ and $J\ll1$, instead, we observe a different behavior for annealed and quenched networks. In the annealed networks there is no phase transition at zero temperature while on quenched network we observe a Griffith phase dominated by extremely rare events and a phase transition at zero temperature.
The Griffiths critical region, nevertheless, is decreasing in size with increasing value of the cutoff $\xi$ of the degree distribution for values of the $\gamma$ exponents $\gamma\leq 3$.
\end{abstract}
\pacs{64.60.aq, 64.70.Tg, 75.10Jm, 89.75.Hc}

\maketitle
\section{Introduction}
The interplay between disorder and superconductivity has attracted the interest of physics in the last decades. Disorder is expected to compete with superconductivity by enhancing the electrical resistivity of a system. In this situation by increasing the random disorder, the system undergoes a superconductor-insulator phase transition \cite{G}.
The theoretical explanation of the interplay between disorder and superconductivity is still a problem of intense debate for granular superconductors. 

Several authors have proposed that high-$T_c$ superconductor are intrinsically inhomogeneous \cite{Kresin, Dagotto, Littlewood,Zaanen2010} however these materials have a multiphase complexity that is difficult to tackle by analytical theoretical models in general.
There is growing interest for a possible optimum inhomogeneity of superconducting cuprates that could enhance the superconducting critical temperature \cite{geballe}. The control of defects and interstitials in heterostructures using new material science technologies can be used to design new granular superconductors with new functionalities \cite{Littlewood}. Recent experiments have provided multiscale imaging of the granular structure of doped cuprate perovskites with a scale invariance that is reflected on the scale-free distribution of oxygen interstitials \cite{Poccia,fratini}. Therefore it is now of high experimental interest the synthesis of novel granular superconductors made of superconducting networks with power law connectivity distribution.
A road map to understand these new possible materials is to construct heterogeneous mean-field models and study the superconductor insulator phase transition in complex scale-free networks.
Recently the superconductor-insulator phase transition has been characterized on Bethe lattices by solving the Random Transverse Ising Model on quenched Bethe lattices \cite{IoffeMezard1, IoffeMezard2,MezardOlga} with the use of advanced quantum cavity methods that allows to go behind the mean-field prediction. These methods have been recently developed to study exact phase diagram of the Random Transverse Ising model \cite{Scardicchio,QCM}. In this paper we use the quantum cavity mapping approximation of these methods,  proposed in \cite{IoffeMezard1,IoffeMezard2,MezardOlga}, that gives a physical interpretation of the phase diagram of the  Random Tranverse Ising Model by mapping the equation determining the phase transition  to a random polymer problem \cite{DS}.
In this paper we extend the results to networks with arbitrary degree distribution.
In particular we focus of scale-free degree distributions mimicking the fractal disorder reported in cuprates.
Already a mean-field calculation of the Random Transverse Ising Model on complex networks has shown that the superconducting critical temperature is strongly enhanced on scale-free degree distribution with power-law exponent $\gamma\leq 3$ \cite{MF}.
This result might explain the experimental findings of Fratini et al. \cite{fratini} and might open a new road map to design new heterogeneous superconductors with enhanced superconducting critical temperature.
On the theoretical side this result is in line with the study of other phase transitions on scale-free network topologies \cite{crit,Dynamics}. In fact it was shown that the classical Ising model \cite{ising,Doro_exp,Bradde}, the percolation transition\cite{Percolation} the epidemic spreading \cite{Epidemic}  change significantly when the second moment $\avg{k(k-1)}$ of the degree distribution diverges. Recently it is becoming clear that these effects of the topology apply also to quantum critical phenomena \cite{sachdev} and are found in the mean-field solution of both the Random Transverse Ising Model \cite{MF} and  the Bose-Hubbard model \cite{BH}.
An open problem, that we approach in this paper is to what extent mean-field results are indicative of the behavior of the critical phenomena in quenched networks.
In particular we focus here on the Random Transverse Ising model and we study the difference between the phase diagram on annealed and quenched scale free networks. While for annealed networks a simple mean-field results is guaranteed to be valid in quenched networks we have to approach the problem by the recently introduced quantum cavity method and by the mapping of this solution to the random polymer problem.
We find that the phase diagram on annealed networks is significantly different from the phase diagram of the quenched networks as long as the second moment of the degree distribution remains finite. In particular we found a phase transition at zero temperature not predicted by the mean-field approach and a replica symmetry broken phase a low temperatures. Instead, as the second moment of the degree distribution diverges $\avg{k(k-1)} \to \infty$ the mean-field predictions approach the quenched solution. 
Therefore we identify the second moment of the degree distribution as a key quantity characterizing the disorder of the topology of the network.
As the second moment of the degree distribution diverges, the superconducting critical temperature of the network diverges both on quenched and on annealed complex networks.

\section{ Random Tranverse Ising model }
We consider a system of  spin variables 
$\sigma_i^z$, for $i=1,\dots,N$, defined on the nodes of a given network with  adjacency matrix ${\bf a}$ such that $a_{ij}=1$ if there is a link between node $i$ and node $j$ otherwise $a_{ij}=0$.
The Random Traverse Ising Model is defined as in \cite{IoffeMezard1,IoffeMezard2,MezardOlga} and is given by the Hamiltonian
 \be 
\hat H=-\frac{J}{2}\sum_{ij}a_{ij}\sigma^z_i \sigma^z_j-\sum_i \epsilon_i \sigma^x_i-h\sum_i \sigma_i^z.
\label{H0}\ee
This Hamiltonian is a simplification respect to the XY model Hamiltonian proposed by Ma and Lee \cite{MaFisher} to describe the superconducting-insulator phase transition but to the leading order the equation for the order parameter is the same as widely discussed in \cite{IoffeMezard1, IoffeMezard2}.
The Hamiltonian describes the superconducting-insulator phase transition as a ferromagnetic spin $1/2$ spin system in a tranverse field.
We propose to use this Hamiltonian to describe  in a granular superconductor in network with heterogeneity of the degree distribution.
 The spins  $\sigma_i$  in Eq. $(\ref{H0})$
 indicate occupied or unoccupied states by a Cooper pair or a localized pair; the parameter $J$ indicates the couplings between neighboring spins, $\epsilon_i$  are quenched values of on-site energy  and $h$ is an external auxiliary field. To mimic the randomness of on-site energy we draw  the variables $\epsilon_i$ from a $\rho(\epsilon)=1/2$ distribution with a finite support $\epsilon\in(-1,1)$.   Finally in  this model the superconducting phase corresponds to the existence of a  spontaneous magnetization in the $z$ direction.  
 
 \section{ Annealed Complex Networks  }
We consider networks of $N$ nodes  $i=1,\ldots,N$. We assign to each node $i$ an hidden variable $\theta_i$ from a $p(\theta)$ distribution indicating the expect number of neighbors of a node. 
An annealed complex network is a network that is dynamically rewired and where the probability to have a link between node $i$ and $j$ is given by  $p_{ij}$  
\be
p_{ij}=P(a_{ij}=1)=\frac{\theta_i \theta_j }{\avg{\theta} N}
 \label{pij2}
\ee

In this ensemble the degree $k_i$ of a node $i$ is a Poisson random variable with expected degree $\overline{k_i}=\theta_i$. 
Therefore we will have 
\bea
\avg{\theta}&=&\overline{\avg{k}}\nonumber \\
\avg{\theta^2}&=&\overline{\avg{k(k-1)}}.
\label{uno}
\eea
where $\avg{\ldots}$ indicates the average over the $N$ nodes of the network and the overline in Eq. $(\ref{uno})$ indicates the average over  time-dependent degrees of the nodes.
 In order to mimic the fractal background present in cuprates \cite{fratini} we assume that the expected degree distribution of the network is given by 
\bea
p(\theta)={\cal N} \theta^{-\gamma}e^{-\theta/\xi}
\label{exp}
\eea
where ${\cal N}$ is a normalization constant and $\xi $ can be modulated by an external parameter such as  doping or strain in cuprates.
 
\section{Solution of the RTIM on  annealed complex networks}
\subsection{Critical temperature}
In oder to study the Random Trasverse Ising Model on annealed complex networks 
we consider  the fully connected Hamiltonian given by 
\be 
\hat {H}=-\frac{J}{2}\sum_{ij}p_{ij}\sigma^z_i \sigma^z_j-\sum_i \epsilon_i \sigma^x_i-h\sum_i \sigma_i^z
\ee
where in order to account for the dynamics nature of the annealed graph we have substituted the adjacency matrix $a_{ij}$ in $H$ with the matrix $p_{ij}$ given by Eq. $(\ref{pij2})$.
In order to evaluate the partition function we  apply the Suzuki-Trotter decomposition in a number $N_s$ of Suzuki-Trotter slices. Therefore we have in the limit $N_s \to \infty$
\bea
\mbox{Tr} e^{-\beta \hat{H}}=\mbox{Tr} \left(e^{-\beta {E}/N_s}e^{-\beta \sum_i \epsilon_i\sigma^x_i/N_s}\right)^{N_s}
\eea
where $E$ is given by 
\bea
E=-\frac{J}{2}\sum_{ij}p_{ij}\sigma^{z}_i \sigma^{z}_j-h\sum_i\sum_{\alpha} \sigma_i^{z}.
\eea
In order to perform this calculation we consider for each spin the sequence ${\underline{\sigma_i^z}}=\{\sigma_i^{z,1},\ldots \sigma_i^{z,N_s}\}$ where each spin $\sigma_i^{z,\alpha}$ represents the spin $i$ in the Suzuki-Trotter slice $\alpha$.
The partition function is then defined as
\bea
Z&=&\sum_{\{\underline{\sigma}_i\}_{i=1,\ldots, N}}\prod_{i=1}^N w(\underline{\sigma_i})e^{\frac{\beta h}{N_s}\sum_{\alpha=1}^{N_s}\sigma_i^{z,\alpha}}  e^{\frac{\beta J}{2 N_s \avg{\theta} N}\sum_{\alpha=1}^{N_s}\sum_{ij}\theta_i \theta_j \sigma_i^{z,\alpha}\sigma_j^{z,\alpha} }
\eea 
where we have indicated with 
\bea
w(\underline{ \sigma_i})=\prod_{\alpha} \langle \sigma_i^{z,\alpha}|e^{\frac{\beta \epsilon_i}{N_s}\sigma^x}|\sigma_i^{z,(\alpha+1)}\rangle.
\eea
In order to disentangle the quadratic terms we use $N_s$ Hubbard-Stratonovich transformations
\bea
Z&=&\left(\frac{\beta N\avg{\theta}}{2\pi N_S}\right)^{N_s/2}\int {\cal D}\underline{S} \exp[-\frac{N\avg{\theta}\beta J}{2N_s}\sum_{\alpha}(S^{\alpha})^2]\nonumber \\
&&\exp[N\int d\theta p(\theta)\frac{1}{2}\int_{-1}^1 d\epsilon \ln \mbox{Tr} \prod_{\alpha}e^{\frac{\beta}{N_s}(h+J \theta S^{\alpha})\sigma^{z}}e^{\frac{\beta}{N_s}\epsilon\sigma^x}], \nonumber
\eea
where ${\cal D}\underline S=\prod_{\alpha=1}^{N_s}d S^{\alpha}$.
The free energy $f=-\frac{1}{\beta}\lim_{N\to \infty } \lim_{N_s \to \infty} \frac{1}{N} \ln Z$ can be evaluated at the stationary saddle point which is cyclically invariant. Therefore we get
\bea
f&=&\mbox{inf}_{S } \frac{J \avg{\theta}}{2}S^2-\frac{1}{\beta}\int d{\theta}p(\theta)\int_0^1 d\epsilon \ln\left(2\cosh\left(\beta \sqrt{(h+JS\theta)^2+\epsilon^2}\right)\right)\nonumber
\eea
where the value of $S$ which minimizes the free energy is given by the saddle point equation
\bea
S&=&\sum_{\theta}\frac{\theta}{\avg{\theta}} p(\theta)\int_0^1 d\epsilon \frac{JS\theta+h}{\sqrt{(JS\theta+h)^2+\epsilon^2}} \tanh(\beta \sqrt{(JS\theta+h)^2+\epsilon^2})
\label{sp}
\eea
Finally the magnetizations along the axis $x$ and $z$ can be calculated by evaluating 
\bea
m_{\theta,\epsilon}^z=\left.\frac{\mbox{Tr }\sigma_{i}^z e^{-\beta \hat{H}}}{Z}\right|_{\theta_i=\theta,\epsilon_i=\epsilon}\nonumber \\
m_{\theta,\epsilon}^x=\left.\frac{\mbox{Tr }\sigma_{i}^x e^{-\beta \hat{H}}}{Z}\right|_{\theta_i=\theta,\epsilon_i=\epsilon}.
\eea 
Performing these calculations we get
\bea 
m_{\theta,\epsilon}^z&=&\frac{JS\theta+h}{\sqrt{(JS\theta+h)^2+\epsilon^2}}\tanh(\beta \sqrt{(JS\theta+h)^2+\epsilon^2})\nonumber \\
m_{\theta,\epsilon}^x&=&\frac{\epsilon}{\sqrt{(JS\theta+h)^2+\epsilon^2}}\tanh(\beta \sqrt{(JS\theta+h)^2+\epsilon^2})\nonumber
\eea
Therefore the magnetizations $m_{\theta,\epsilon}^z$ and $m_{\theta,\epsilon}^x$ depend on the value  $\theta$ of the expected degree of the node and on the on-site energy $\epsilon$.

The order parameter for the superconducting-insulator phase transition is $S$ given by Eq. $(\ref{sp})$.
From the self-consistent equation determining the order parameter for the transition it is immediate to show that the  superconducting-insulator  phase transition occurs for $h=0$ at 
\be
1=J\frac{\avg{\theta^2}}{\avg{\theta} }\int_0^1 d\epsilon \frac{\tanh(\beta \epsilon)}{\epsilon}=\frac{J}{J_c(\beta)}
\ee
which implies that for $\avg{\theta^2}\to \infty$ then $\beta\to 0$  for any fixed value of the coupling $J>0$ ,  and the critical temperature for the paramagnetic ferromagnetic phase transition $T_c$ diverges.
This implies that on annealed scale free networks with $p(\theta)\propto \theta^{-\gamma}$ and $\gamma\leq 3$   the Random Tranverse Ising model  is always in the superconducting phase.

If we consider the special case of an expected degree distribution given by Eq. $(\ref{exp})$ and  $\gamma\leq 3$ we have that $\frac{\avg{\theta^2}}{\avg{\theta}}$ diverges in the limit $\xi\to \infty$.
If we take $J\frac{\avg{\theta^2}}{\avg{\theta}}\gg 1$ which can be achieved by keeping $J$ constant and we go in the limit $\xi\to \infty$  with $\gamma\leq 3$ we find that the critical temperature for the superconductor-insulator transition diverges as 
\begin{equation}
T_c\propto J\frac{\avg{\theta^2}}{\avg{\theta}}\propto \left\{\begin{array}{cc}\ln \xi & \mbox{if } \gamma=3\nonumber \\
\xi^{3-\gamma} &\mbox{if } \gamma<3\end{array}\right.
\end{equation}
Therefore on complex topologies when the expected degree distribution is given by Eq. $(\ref{exp})$ and if $\gamma\leq3$ we observe an enhancement of the superconducting temperature with  increasing  values of the cutoff  $\xi$.
On the contrary for $J\frac{\avg{\theta^2}}{\avg{\theta}}\ll1$ we have
\begin{equation}
T_c=\frac{4e^C}{\pi} \exp\left[-\frac{\avg{\theta}}{J \avg{\theta^2}}\right]
\label{tcz}
\end{equation}
where $C\sim 0.577$ is the Euler number.
Therefore in the annealed network model there is no phase transition at zero temperature. In fact from Eq. $(\ref{tcz})$ when $\avg{\theta^2}$ is finite  we have that $T_c=0$ only for $J=0$.
This phenomena strongly depends on the assumption that the network is annealed as it was found in \cite{IoffeMezard1,IoffeMezard2} where the critical behavior of the Random Tranverse Ising model was studied on a Bethe lattice.
The critical indices of the phase transition will be given by the heterogeneous mean-field \cite{Doro_exp} approach to phase transition as long as the cutoff $\xi=\infty$ and will depend non-trivially from the value of the power-law exponent of the expected degree distribution $\gamma$. The critical index of the susceptibility will be the only exception and remain mean-field as long as we don't include dependence of the network on the embedding space \cite{Bradde}. 

\section{Quenched network}
In this section we introduce quenched networks that have a structure that does not change in time. In the subsequent section we will characterize the Random Transverse Ising model on quenched networks and we compare the results with the heterogeneous mean-field results obtained in the previous sections.
Therefore we consider a quenched network of $N$ nodes $i=1,2\ldots, N$ and degree distribution $p(k)$. We indicate by $N(i)$ the set of nodes neighbors of node $i$. We assume that the network is random meaning that the probability to reach a node of degree $k$ by following a random link is given by $k p(k)/\avg{k}$. We assume that the distribution of quenched local energies $\epsilon_i$ is given by $\rho(\epsilon)=1/2$ with $\epsilon\in(-1,1)$.
We will consider in particular the case in which the degree distribution $p(k)$ of the network is scale-free, with power-law exponent $\gamma$, and regularized by an exponential cutoff $\xi$, i.e. we assume
\begin{equation}
p(k)={\cal N} k^{-\gamma} e^{-k/\xi}.
\label{pke}
\end{equation}
where ${\cal N}$ is a normalization constant.

\section{Solution on a quenched network}
In quenched networks critical phenomena might show a different phase diagram than on annealed networks.
In particular this is true for the Random Transverse Ising model that on quenched Bethe lattices has a zero temperature phase transition which is absent on the corresponding annealed network as discussed in \cite{IoffeMezard1,IoffeMezard2,MezardOlga}.
In order to study critical phenomena on quenched networks with a locally tree like structure (vanishing clustering coefficient)  the theory of the cavity method has been developed and applied to a large variety of classical critical phenomena on networks. Recently this approach has been extended to quantum critical phenomena without disorder using the Suzuki-Trotter formalism finding exact results \cite{QCM,Scardicchio}. In this paper we take  a different approach by following a recent method \cite{IoffeMezard1,IoffeMezard2,MezardOlga} proposed to study the  Random Transverse Ising model  on Bethe lattices. This approach has the advantage that it can be applied to quantum critical phenomena with disorder and that it is analytically tractable. Moreover, as we will see, the solution of the problem can be mapped to a directed polymer problem shedding light on the nature of the different phases on the critical dynamics.
\begin{figure}
\begin{center}
\includegraphics[width=.6\columnwidth]{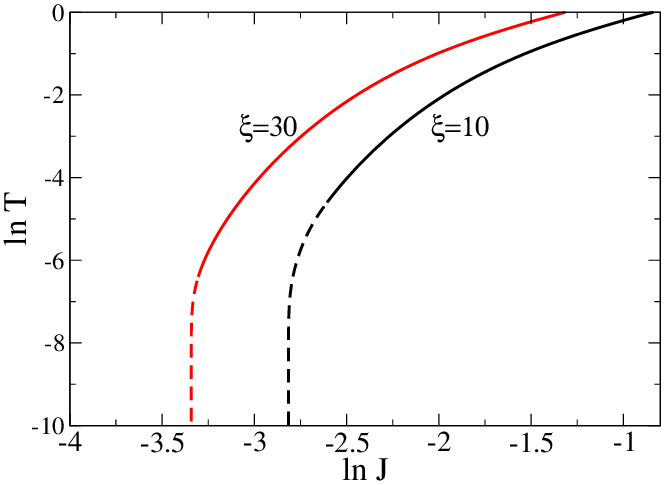}
\end{center}
\caption{(Color online)    Critical line $T_c=T_c(J)$ for the Random Tranverse Ising Model on a complex network with power law exponent $\gamma=2.5$ and different values of the cutoff $\xi$. The dashed lines indicate the onset of the Griffith (replica-symmetry broken) phase. This picture show that the Griffith phase occupies a reduced region of phase space if the cutoff $\xi$ and the branching ratio $\frac{\avg{k(k-1}}{\avg{k}}$ diverges. }
\label{RS_RSB}
\end{figure}

\begin{figure}
\begin{center}
\includegraphics[width=.6\columnwidth]{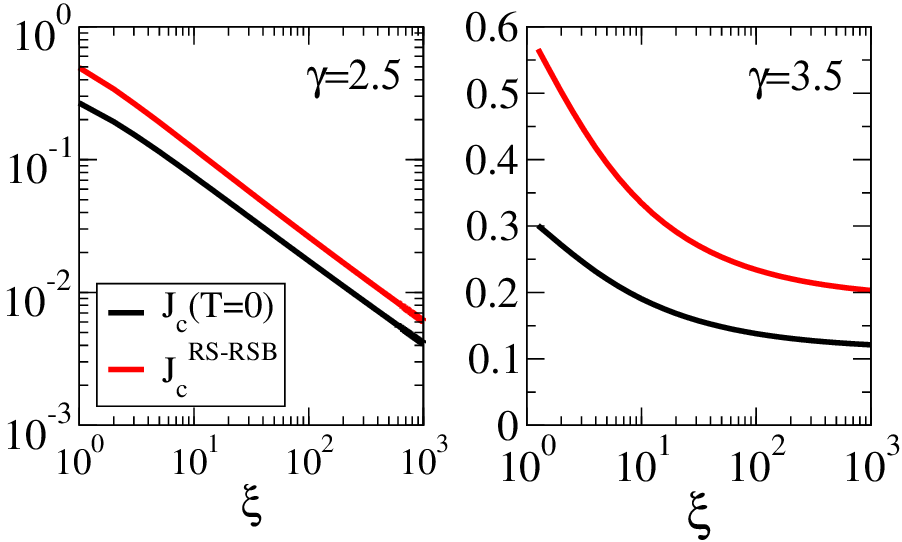}
\end{center}
\caption{(Color online)    Critical value $J_c(T=0)$ for the insulator-superconductor phase transition and critical value $J_c^{RS+RSB}$ for the replica-symmetric and replica symmetry broken phase. For $\gamma=2.5$ the Griffith phase occupies a reduced fraction of the phase space as the exponential cutoff $\xi$ diverges while for $\gamma=3.5$ the Griffith phase survives also in the $\xi\to \infty$ limit.}
\label{Jc}
\end{figure}
\begin{figure}
\begin{center}
\includegraphics[width=.6\columnwidth]{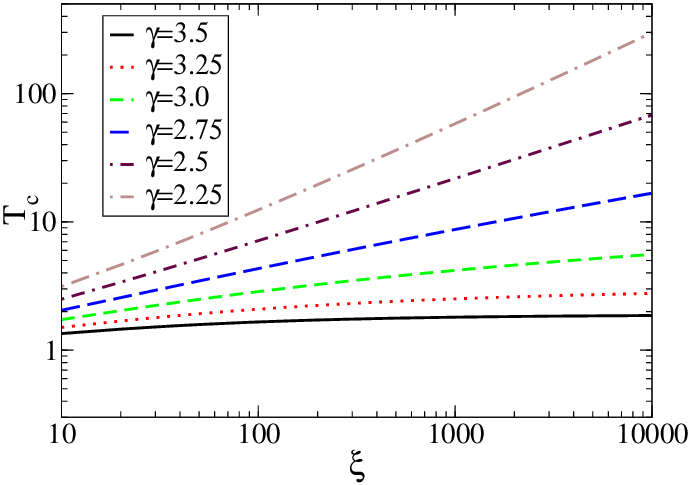}
\end{center}
\caption{(Color online)  The critical temperature of the superconductor insulator phase transition as a function of the cutoff $\xi$ when the coupling constant $J=1$ is kept constant. As $\gamma\leq3$ the critical temperature for superconductivity is strongly enhanced by the scale-free topology of the underlying network.}
\label{Tc_xi}
\end{figure}
\subsection{Cavity mapping}
In the cavity model, the cavity graphs is considered. The cavity graph is the graph centred on a node $i$ when one of its neighbor $j\in N(i)$ is removed. It is therefore assumed that all the other neighbors $\alpha \in N(i) \setminus j$ of node $i$ are uncorrelated. On each of these nodes $\alpha$ it is assumed that an efficient cavity field $B_{\alpha,i}$ is acting such that the local Hamiltonian acting on the cavity graph centred in node $i$ is given by 
\begin{equation}
H_{i,j}^{cav}=-\epsilon_i \sigma_i^{x}-\sum_{\alpha \in N(i) \setminus j}[\epsilon_{\alpha}\sigma_{\alpha}^x+B_{\alpha, i}\sigma_{\alpha}^z+J\sigma_i^z\sigma_{\alpha}^z].
\end{equation}
where we assume here and in the following that the external field $h=0$.
It was recently shown that the cavity-mean-field method  is a very useful tool to study the Random Tranverse Ising model \cite{IoffeMezard1,IoffeMezard2,MezardOlga}. In fact on the same time it provides an analytical approach to the solution of the Hamiltonian, and it  provides a physical interpretation of the phenomena occurring at low temperatures. Finally this approximation is able  to reproduce with good accuracy the phase diagram of the Random Transverse Ising Model.
In the  cavity mean-field approximation we replace the dynamical spin variables $\sigma_{\alpha}^z$ in $H_{i, j}^{cav}$ by their averages  $\avg{\sigma_{\alpha}^z}$ with values
\begin{equation}\avg{\sigma_{\alpha}^z}=\frac{B_{\alpha, i}}{\sqrt{B_{\alpha, i }^2+\epsilon_{\alpha}^2}}\tanh{\beta \sqrt{B_{\alpha, i}^2+\epsilon_{\alpha}^2}}. \end{equation} 
Therefore we  assume that the cavity Hamiltonian acting on spin $i$  in absence of spin $j$ can be approximated by
\begin{equation}
H^{cav−MF}_{i,j}=-\epsilon_i\sigma^x_i-J\sigma^z_i\sum_{\alpha\in N(i)\setminus j}\avg{\sigma^z_{\alpha}}
\end{equation}
which implies that $B_{i, j}=J\sum_{\alpha \in N(i)\setminus j}\avg{\sigma_{\alpha}^z}$ and therefore
\begin{equation}
B_{i, j}=J\sum_{\alpha\in N(i)\setminus j}\frac{B_{\alpha, i}}{\sqrt{B_{\alpha, i }^2+\epsilon_{\alpha}^2}}\tanh{\beta \sqrt{B_{\alpha, i}^2+\epsilon_{\alpha}^2}}
\label{recursive}
\end{equation}
This recursion induces a self-consistent equation on the distribution $P_k(B)$ of the cavity fields $B$ for nodes of degree $k$
\bea
P_k(B)&=&\prod_{\alpha}\sum_{k_{\alpha}}\frac{p(k_{\alpha})k_{\alpha}}{\avg{k}}\int \prod_{\alpha}[d B_{\alpha} d \epsilon_{\alpha}  P_{k_{\alpha}}(B_{\alpha})\rho(\epsilon_{\alpha})]\nonumber \\&&\delta\left(B-J\sum_{\alpha=1}^{k-1}\frac{B_{\alpha}}{\sqrt{B_{\alpha^2}+\epsilon_{\alpha}^2}}\tanh{\beta\sqrt{B_{\alpha^2}+\epsilon_{\alpha}^2}}\right).
\label{pkB}
\eea
\subsection{Mapping of the problem to a direct polymer}
In order to characterize the critical behavior of the Random Transverse Ising Model we can imagine to iterate the relation Eq. $(\ref{recursive})$ on the complex network for $L \gg1$ times. For $L$ finite and $N \to \infty$ the corresponding graph is locally a tree with branching ratio
\begin{equation}
K=\frac{\avg{k(k-1)}}{\avg{k}}.
\end{equation}
The cavity field at the root of this tree is a function of the $K^L$ cavity fields on the boundary. If we assume that infinitesimal cavity fields $B\ll1 $ are acting on the boundary nodes the cavity field $B_0$ at the root of the tree is given by
\begin{equation}
\frac{B_0}{B}=\sum_{{\cal P}} \prod_{\alpha \in {\cal P}} J \frac{\tanh{\beta \epsilon_{\alpha}}}{\epsilon_{\alpha}}=\Xi
\label{polymer}
\end{equation}
where ${\cal P}$ indicates the path on the tree from the root to the leaf of the tree, and  the product $\prod_{\alpha \in {\cal P}}$ is over all node along the path ${\cal P}$.
The critical point of the dynamics is given by the parameters for which $\frac{1}{L}\ln\Xi=0$ indicating when the cavity field propagates over large distances on the network.
The expression Eq. $(\ref{polymer})$ allows for a mapping of our problem to the directed polymer (DP) problem. In particular the function $\Xi $ is the partition function of the directed polymer where the energy on each edge is given by $e^{-E_{\alpha}}=J\frac{\tanh{\beta \epsilon_{\alpha}}}{\epsilon_{\alpha}}$ and the temperature is set equal to one.
This problem has been studied by Derrida and Spohn \cite{DS}. In the directed polymer problem there are two regimes:
\begin{itemize}
\item
The replica symmetric (RS) regime  in which the measure defined in Eq. $(\ref{polymer})$ is more or less evenly distributed among the paths
\item
The replica symmetry broken (RSB) regime in which the measure defined in Eq. $(\ref{polymer})$ condenses on a small number of paths.
\end{itemize}
In order to find the phase transition between the two phases we need to study the behavior of  $\ln \Xi$ over a typical sample of the disorder. This is done by evaluating the average over the quenched variables of $\overline{\ln \Xi}$.
This computation can be done by making use of the replica method \cite{MPV,IoffeMezard1,IoffeMezard2,MezardOlga} by computing
\begin{equation}
\overline{\ln\Xi}=\lim_{n\to 0} \frac{\overline{\Xi^n}-1}{n}
\end{equation}
where the replicated partition function is a sum over $n$ paths
\begin{equation}
\overline{\Xi^n}=\sum_{{\cal P}_1,{\cal P}_2,\ldots, {\cal P}_n}\prod_{\alpha} \overline{\left(J \frac{\tanh{\beta \epsilon_{\alpha}}}{\epsilon_{\alpha}}\right)^{r_{\alpha}}}
\end{equation}
where the weight of each node $\alpha$ depends on the number of paths which go though this node.
In the RS solution we assume that all the paths are non-overlapping and independent, 
therefore
\begin{equation}
\frac{1}{L}\overline{\ln \Xi}=\ln J+\ln\left[\frac{\avg{k(k-1)}}{\avg{k}} \int_0^1 d\epsilon \frac{\tanh{\beta \epsilon}}{\epsilon}\right]=\ln(x)+f(1)
\end{equation}
In the RSB solution we assume instead that the $n $ paths consist of groups of $n/x$ identical paths, and therefore $r_{\alpha}=x$. With this assumption we get
\begin{equation}
\frac{1}{L}\overline{\ln \Xi}=\ln J+\frac{1}{x}\ln\left[\frac{\avg{k(k-1)}}{\avg{k}} \int_0^1 d\epsilon  \left(\frac{\tanh{\beta \epsilon}}{\epsilon}\right)^x\right]=\ln J +f(x)
\end{equation}
where 
\begin{equation}
f(x)=\frac{1}{x}\ln \left[\frac{\avg{k(k-1)}}{\avg{k}}\int_0^1 d\epsilon \left(\frac{\tanh{\beta \epsilon}}{\epsilon}\right)^x\right].
\label{f}
\end{equation}
The value of $x$ is found by extremizing the function $f(x)$ respect to $x$ with $x\in [0,1]$.
In particular by minimizing $f(x)$ as a function of $x$ one find that for $\beta>\beta_{RSB}$ the solution is replica symmetry broken with $x=m<1$ and for $\beta>\beta_{RSB}$ the function is minimized at the boundary $m=1$ and the solution is replica symmetric.

\subsection{Phase diagram}
The critical behavior of the Random Tranverse Ising Model is found when $\frac{1}{L}\overline{\ln \Xi}=0$ and when $B_0/B$ the cavity field at the boundary propagates at long distances on the network under the iteraction Eq. $(\ref{recursive})$. 
Therefore we have to distinguish between  two regimes:
\begin{itemize}
\item
The replica symmmetric (RS) regime for $\beta<\beta_{RSB}$.\\
In this regime we have that the critical line is dictated by the relation 
$\frac{1}{L}\overline{\ln \Xi}=f(1)+\ln(J)=0$, therefore the critical points are solution to the equation
\begin{equation}
J \frac{\avg{k(k-1)}}{\avg{k}}\int_0^1 d\epsilon  \frac{\tanh{\beta \epsilon}}{\epsilon}=1
\end{equation}
In the limit $J \frac{\avg{k(k-1)}}{\avg{k}}\gg1$
 we get
 \begin{equation}
 T_c=J\frac{\avg{k(k-1)}}{\avg{k}}\propto \left\{\begin{array}{cc}\ln \xi & \mbox{if } \gamma=3\nonumber \\
\xi^{3-\gamma} &\mbox{if } \gamma<3\end{array}\right.
 \end{equation}
where the last relation is valid  if $p(k)={\cal N} k^{-\gamma}e^{-k/\xi}$ .
\item
The replica symmmetry broken (RSB) regime for $\beta>\beta_{RSB}$.\\
In this regime we have that the critical line is dictated by the relation 
$\frac{1}{L}\overline{\ln \Xi}=\ln(J)+f(m)=0$ and $f^{\prime}(m)=0$, therefore the critical points are solution to the system of equations
\begin{eqnarray}
J^m \frac{\avg{k(k-1)}}{\avg{k}}\int_0^1 d\epsilon  \left(\frac{\tanh{\beta \epsilon}}{\epsilon}\right)^m=1 \nonumber \\
\int_0^1 d\epsilon  \left(\frac{\tanh{\beta \epsilon}}{\epsilon}\right)^m \ln\left[J\frac{\tanh{\beta \epsilon}}{\epsilon}\right]=0
\label{rs-rsb}
\end{eqnarray}
\end{itemize}

\subsubsection{Phase diagram at $T=0$}
It is interesting to study the critical point at $T=0$. In particular we observe relevant deviations of the phase diagram on a quenched network respect to the behavior on a annealed network when the critical point is predicted to be at $J=0$. 
The function $f(m)$ defined in Eq. $(\ref{f})$ take the simple form
\begin{equation}
f(m)=\frac{1}{m}\ln\left[\frac{\avg{k(k-1)}}{\avg{k}}\frac{1}{1-m}\right].
\end{equation}
At $T=0$ we can study the critical point of the transition by imposing $f^{\prime}(m)=0$ and $\frac{1}{L}\overline{\ln \Xi}=\ln(J)+f(m)=0$. By studying these equation we get the critical point
\begin{equation}
\frac{1}{J_c}=e \frac{\avg{k(k-1)}}{\avg{k}}\ln\left(e\frac{\avg{k(k-1)}}{\avg{k}}\right).
\end{equation}
Therefore on any given network with finite branching ratio there is a quantum phase transition at $T=0$.
In the case in which $p(k)={\cal N} k^{-\gamma}e^{-k/\xi}$ and $\gamma\leq 3$ we observe for large value of the exponential cutoff $\xi$ that $J_c$ goes to zeros as 
\begin{equation}
J_c\propto \left\{ \begin{array}{cc}\frac{1}{\xi^{3-\gamma}\ln(\xi)} & \mbox{if } \gamma<3\nonumber\\
\frac{1}{\ln\xi \ln(\ln\xi)} &\mbox{if } \gamma=3 \end{array}\right.
\end{equation}
Therefore in the $\xi\to \infty$ limit, as long as the power-law exponent $\gamma$ is $\gamma\leq3$ we recover the mean field result that there is no phase transition at zero temperature.
\subsubsection{Phase diagram for $T>0$:  RS-RSB Phase Transition}
It is interesting to find the critical point $T_c=T_{RSB}\ll1$ where we have the transition between the replica symmetric phase and the replica symmetry broken phase.
This point is found by solving the equations Eqs. $(\ref{rs-rsb})$ for $m=1$ which we rewrite here for convenience.
\begin{eqnarray}
J \frac{\avg{k(k-1)}}{\avg{k}}\int_0^1 d\epsilon \left(\frac{\tanh{\beta \epsilon}}{\epsilon}\right)=1 \nonumber \\
\int_0^1 d\epsilon \left(\frac{\tanh{\beta \epsilon}}{\epsilon}\right) \ln\left[J\frac{\tanh{\beta \epsilon}}{\epsilon}\right]=0
\label{RS_RSB_P}
\end{eqnarray}
We can find the critical point by expanding the equation for $J\frac{\avg{k(k-1)}}{\avg{k}}\ll1$ and $T\ll 1$ finding
\begin{eqnarray}
\frac{1}{J_c^{RS-RSB}}=2 \frac{\avg{k(k-1)}}{\avg{k}}\ln\left(2 \frac{\avg{k(k-1)}}{\avg{k}}\right)\nonumber \\
T_c^{RS-RSB}=\frac{e^C}{\pi}\left[\frac{\avg{k(k-1)}}{\avg{k}}\ln\left(2\frac{\avg{k(k-1)}}{\avg{k}}\right)\right]^{-2}
\end{eqnarray}
where $C$ is the Euler constant $C=0.577..$.
Therefore the critical temperature for the onset of the RSB phase for degree distributions given by Eq. $(\ref{pke})$ and $\gamma\leq 3$ goes to zero as the exponential cutoff $\xi$ diverges, and we have
\begin{equation}
T_c^{RS_RSB}\propto \left\{ \begin{array}{cc}\frac{1}{\xi^{2(3-\gamma)}\ln^2(\xi)} & \mbox{if } \gamma<3\nonumber\\
\frac{1}{[\ln\xi \ln(\ln\xi)]^2} &\mbox{if } \gamma=3 \end{array}\right..
\end{equation}
In Figure $\ref{RS_RSB}$ we plot the critical temperature for the phase transition between the superconducting and insulator phase for small value of the temperature $T$ and coupling $J$.
For temperature $T<T_{RS-RSB}$ the critical line is dictated by the Replica-Symmetry broken equations Eq. $(\ref{RS_RSB_P})$.
By plotting $T_c=T_c(J)$ for networks with different value of the branching ratio $K=\frac{\avg{k(k-1)}}{\avg{k}}$ we show that as  $K$ grows the replica symmetry broken phase shrinks.
In order to show how severe it is this effect in Figure $\ref{Jc}$ we show the critical coupling constant $J_c^{RS_RSB}$ and the critical coupling constant at $T=0$ $J_c=J_c(T=0)$ as a function of the cutoff and the power-law exponent $\gamma$. We show that as the power-law exponent $\gamma\leq 3$ the branching ratio $K$ diverges with diverging values of the cutoff $\xi$ and in the limit $\xi\to \infty$ we recover the mean-field results.
Moreover as we show in Figure $\ref{Tc_xi}$, as the coupling is fixed  and $J\frac{\avg{k(k-1)}}{\avg{k}}\gg1$, the critical temperature for the superconductor-insulator phase transition diverges
with diverging values of the branching ration $K=\frac{\avg{k(k-1)}}{\avg{k}}$.
Finally the ciritical indices of the phase transition will depend on the value of the exponential cutoff $\xi$ and the power-law exponent $\gamma$ of the degree distribution.
Moreover, following similar arguments used in \cite{IoffeMezard2} we cha show that the replica symmetric phase is a Griffith phase.

\section{Conclusions}
In conclusion, we have studied the Random Transverse Ising model on networks with arbitrary degree distribution as a paradigm to study superconductor-insulator phase transitions.
We have studied the model both on annealed and quenched networks using mean-field and quantum cavity method. In particular we have chosen a recently proposed approximation of the quantum cavity method that allows for a full analytical treatment of the problem while characterizing the phase diagram going beyond the mean-field treatment.
This method is based on a mapping between the cavity equation and the random polymer problem in quenched media.
We have therefore characterized fully the differences between the phase diagram of the Random Transverse Ising model defined on annealead and quenched networks.
The model shows a significant dependence on the second moment of the degree distribution $\avg{k(k-1)}$. In particular the superconducting critical temperature is enhanced in networks with greater second moment of the degree distribution.
Moreover in the limit $\avg{k(k-1)}\to \infty$ the phase diagram of the quenched networks is well approximated by the phase diagram of the annealed networks.

\acknowledgments
We thank Marc M\'ezard for stimulating discussions and for hospitality in LPTMS where this work started.

\end{document}